\documentclass{emulateapj}
\usepackage{lscape}
\usepackage{amsmath}

\newcommand{\lya}{Ly$\alpha$ }

\newcommand{\zzzp}{$z\sim3$}
\newcommand{\zzz}{$z\sim3$ }

\newcommand{\kpc}{$h^{-1}$kpc }

\shorttitle{Broadband \zzz galaxy segregation} 
\shortauthors{Cooke} 

\begin{document}

\title{Broadband Imaging Segregation of \zzz \lya Emitting and \lya
Absorbing Galaxies}
 
\author{Jeff Cooke\altaffilmark{1}\altaffilmark{2}} 
\affil{The Center for Cosmology and the Department of Physics \&
Astronomy, The University of California, Irvine, Irvine, CA,
92697-4575}

\email{cooke@uci.edu}
\altaffiltext{1}{California Institute of Technology, MC 249-17, 1200
E. California Blvd., Pasadena, CA 91125} 
\altaffiltext{2}{Gary McCue Postdoctoral Fellow}


\begin{abstract}
The spectral properties of Lyman break galaxies (LBGs) offer a means
to isolate pure samples displaying either dominant \lya in absorption
or \lya in emission using broadband information alone.  We present
criteria developed using a large \zzz LBG spectroscopic sample from
the literature that enables large numbers of each spectral type to be
gathered in photometric data, providing good statistics for multiple
applications.  In addition, we find that the truncated faint, blue-end
tail of \zzz LBG population overlaps and leads directly into an
expected \lya emitter (LAE) population.  As a result, we present
simple criteria to cleanly select large numbers of \zzz LAEs in deep
broadband surveys.  We present the spectroscopic results of $32$
$r'\lesssim25.5$ LBGs and $r'\lesssim27.0$ LAEs at \zzz pre-selected
in the Canada-France-Hawaii Telescope Legacy Survey that confirm these
criteria.
\end{abstract}



\section{INTRODUCTION}\label{intro}

The Lyman break galaxies \citep[LBGs;][]{s96} are a high-redshift
population of star forming galaxies selected by their rest-frame
ultraviolet colors.  LBGs at \zzz are faint ($L*$ corresponds to
m$_R\sim24.5$ at \zzzp) and require long integrations using 8 m class
telescopes to obtain spectral information having a signal-to-noise
ratio (S/N) of a few.  Nevertheless, $>1500$ \zzz spectra have been
obtained and studied \citep[e.g.,][hereafter
CCS03]{lefevre05,c05,s03}. \lya is the most prominent feature in LBG
spectra and has been shown to be indicative of other spectral
properties, such as interstellar medium absorption line strength and
continuum profile \citep[][hereafter AES03]{aes03}.

Inspection of the spectroscopic samples gathered to date show that
$\sim50$\% of LBGs exhibit dominant (or net) \lya in absorption, with
the remaining exhibiting dominant \lya in emission \citep[e.g.,
AES03][]{c05}.  A recent investigation of close and interacting pairs
\citep{c09a} finds evidence of an overabundance of \zzz LBGs
exhibiting \lya emission in close pairs; all LBGs with projected
physical separations $\lesssim15$ \kpc display \lya in emission.  To
properly explore this relationship, along with other spectroscopic
trends with spatial distribution, large samples that reflect the
spectral properties of LBGs are necessary.  However, the conventional
means of multi-object spectroscopy is inefficient in acquiring the
spectra of galaxies closely spaced on the sky because of inherent
mechanical constraints \citep[see][]{c09a}.  As a result,
comprehensive LBG spectroscopic surveys are difficult and
time intensive.

High-redshift star-forming systems with continua typically too faint
for spectroscopic follow-up, but with prominent \lya emission
\citep{cowie98,hu98}, have been detected via targeted spectroscopic
and narrowband surveys
\citep[e.g.,][]{dawson04,ouchi05,venemans05,gaw06}.  Because of the
faint continua of these \lya emitters (LAEs), and the difficulties
involved in their acquisition, LAE surveys have been limited to narrow
redshift ranges and typically clustered fields.  As a result, fewer
total spectra have been compiled relative to LBGs, with the details of
the process, or processes, that contribute to the generation and
escape of \lya emission, less certain.  In addition, the
data-gathering techniques have provided only a few cases to date where
the mass of LAEs from clustering can be reasonably inferred
\citep{ouchi03,shimasaku04,kovac07,gaw07}.

LBGs at \zzz that are dominated by \lya in absorption separate
sufficiently in color--magnitude space from those dominated by \lya in
emission to enable a simple means using broadband imaging to isolate
subsets with desired \lya features and associated spectral properties.
Furthermore, we find that deep broadband criteria can select clean, EW
unbiased samples of \zzz LAEs with photometrically detectable
continua.  As a result, very large numbers \zzz systems over defined
volumes with desired spectral properties can be efficiently obtained
for statistical study, complementing the limitations of narrowband
surveys and extensive deep spectroscopic campaigns.  We describe the
observations used in this work in Section~\ref{data}.  In
Section~\ref{analysis}, we present the behavior of LBG spectroscopic
subsets, the criteria for LBG spectral-type and LAE selection, and the
results of our spectroscopic test of the criteria.  Finally, we
provide a conclusion in Section~\ref{disc}.  Magnitudes presented here
are in the AB \citep{f96} magnitude system.

\begin{figure*}
\begin{center}
\scalebox{0.35}[0.33]{\rotatebox{90}{\includegraphics{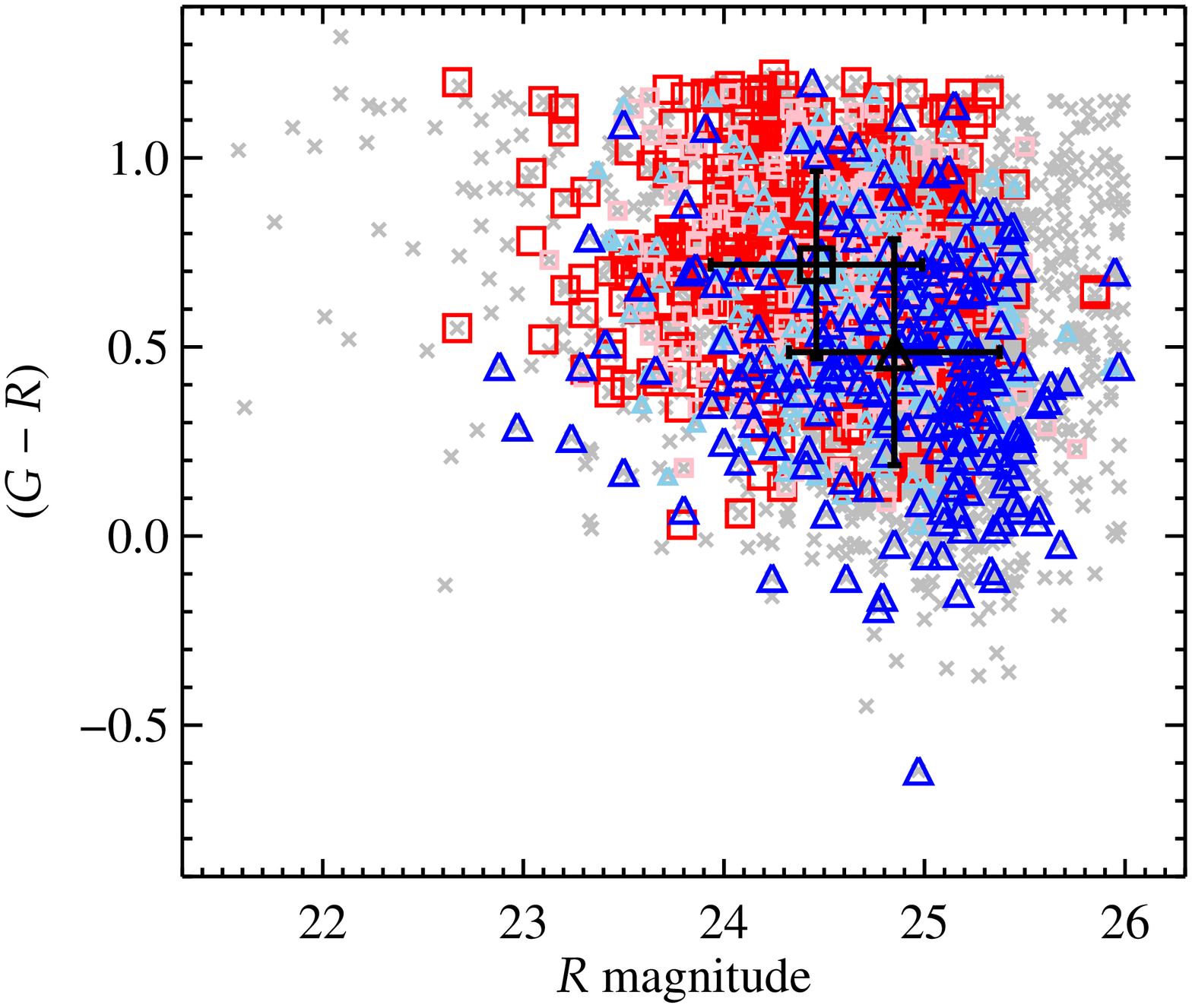}}}
\scalebox{0.35}[0.33]{\rotatebox{90}{\includegraphics{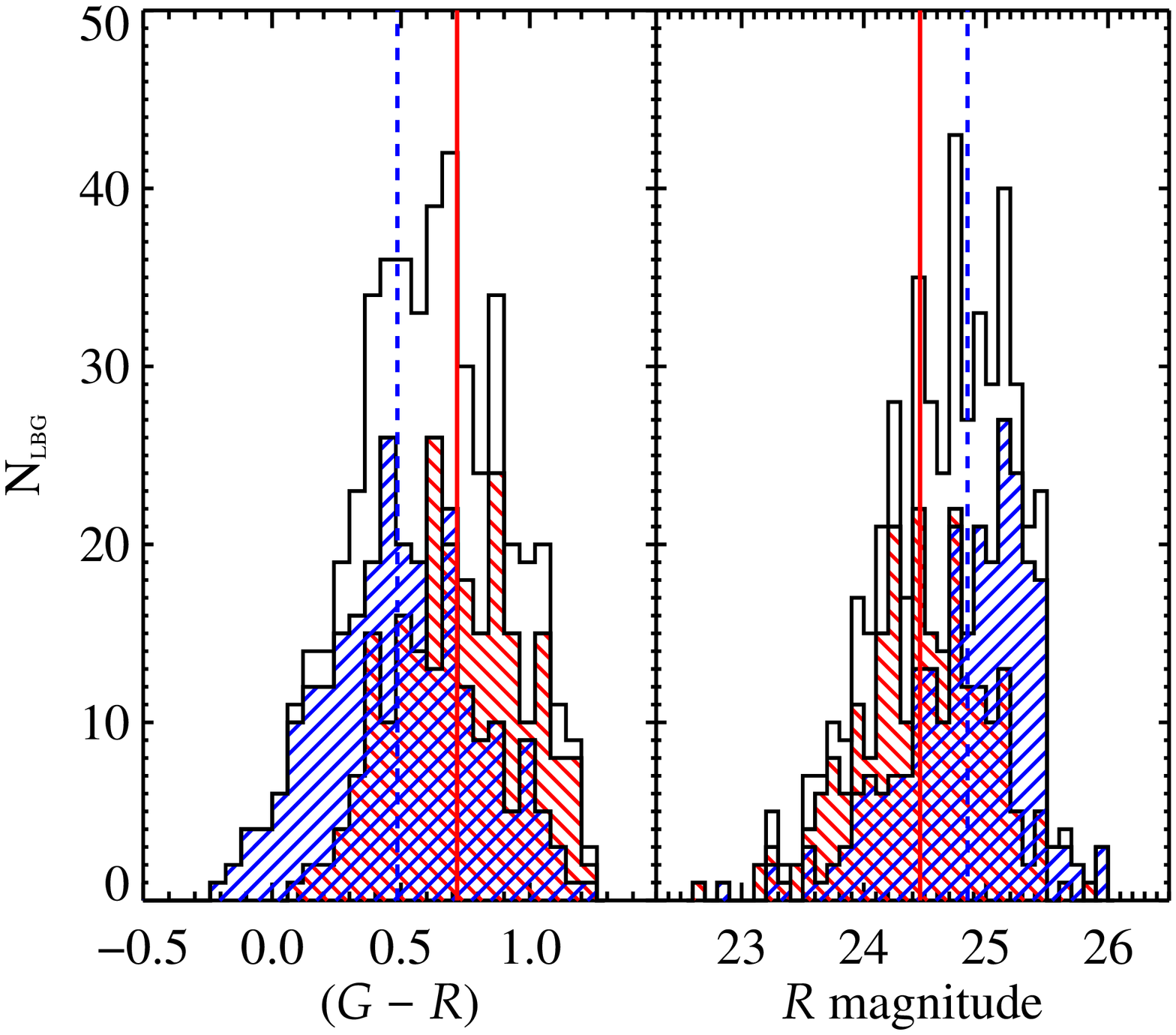}}}
\caption
{\small Left panel: color-magnitude diagram of \zzz LBGs in the
  \citet{s03} sample.  Photometrically selected galaxies are plotted
  as gray crosses, whereas spectroscopically confirmed \zzz LBGs are
  the larger (colored) symbols coded by their \lya EW and represent
  the four quartiles as defined in \citet{aes03}. Specifically, group
  1 -- large {\it (gray/red)} squares, group 2 -- small {\it
  (gray/pink)} squares, group 3 -- small {\it (black/light blue)}
  triangles, and group 4 -- large {\it (black/blue)} triangles.  The
  large square and triangle and associated error bars represent the
  mean and $1\sigma$ standard deviation for the distributions of LBGs
  with net \lya in absorption (aLBGs) and net \lya in emission
  (eLBGs), respectively.  Right two panels: histograms of the aLBG
  ({\it (gray/red)}, back-hatched) and eLBG ({\it (black/blue)},
  forward-hatched) ($G-\cal{R}$) color and $\cal{R}$ mag distributions
  (see the text).  The ($G-\cal{R}$) color-selection criteria omit the
  far red tail of the aLBG distribution in an effort to avoid
  low-redshift interlopers.  The spectroscopic-justified magnitude
  truncation ($\cal{R}$ $\le 25.5$) of eLBGs suggests that a
  continuation of this distribution probes a \lya emitter
  population. }
\label{color}
\end{center}
\end{figure*}


\section{OBSERVATIONS}\label{data}

We use the publicly available data set of CCS03
\footnote{http://vizier.cfa.harvard.edu/viz-bin/VizieR?-source=J/ApJ/592/728/}
and related accessible files to quantify the spectroscopic properties
of \zzz LBGs and establish the $U_n G \cal{R}$ spectral-type selection
criteria.  This survey consists of $\sim2500$ $U_n G \cal{R}$ selected
and $\sim800$ spectroscopically confirmed LBGs from 17 separate fields
that effectively sample the \zzz LBG population as a whole.  As a key
component to this work, we use the \lya EW measurements of AES03 for
$775$ LBG spectra of the CCS03 sample (A. E. Shapley, 2009, private
communication).
 
In Section~\ref{test}, we test the criteria presented below using the
spectra of $32$ $u^*gi$-selected \zzz LBGs in the Canada France Hawaii
Telescope Legacy Survey (CFHTLS) Deep field ``D4''
\footnote{http://www1.cadc-ccda.hia-iha.nrc-cnrc.gc.ca/cadcbin/cfht/wdbi.cgi/cfht/tpx\_fields/form}.
We $r$-band select \zzz LBGs from stacked images that combine four
years of high-quality observations that reach limiting magnitudes of
$u^*\sim27.5,g\sim27.5,r\sim27.0,$ and $i\sim26.7$.  The stacked
$r$-band images probe $\gtrsim1.0-1.5$ mag deeper than the
$\cal{R}$-band images of CCS03 and thereby test a fainter regime.  The
\zzz LBG criterion $(u^*-g)\ge1.25(g-i)$ for the CFHT Megacam filters
imposes a weaker restriction on the $u^*$-band depth as compared to
the $(U_n-G)\ge(G-$ $\cal{R}$$)+1.0$ criteria of CCS03.  This
minimizes the introduction of a bias in $r>25.5$ LBG selection based
on their $(g-i)$ colors.  For further information on image stacking
and color-selection spectroscopic tests, see \citet[][Supplementary
Information]{c09b}.

\begin{figure}
\begin{center}
\scalebox{0.35}[0.33]{\rotatebox{90}{\includegraphics{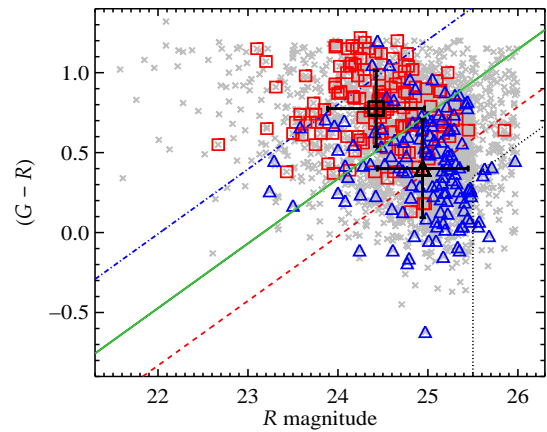}}}
\caption
{\small Similar to Figure~\ref{color}, but for LBGs that exhibit the
  opposite extremes in \lya features and associated spectral profiles.
  Squares denote the subset of $150$ LBGs that display the strongest
  \lya absorption and triangles denote the subset of $150$ LBGs that
  display the strongest \lya emission. The solid (green) line denotes
  our primary statistical cut of the two distributions and is the
  basis for subsequent selection criteria.  The dashed (red) and
  dot-dashed (blue) lines indicate a ($G-\cal{R}$) distribution
  $1.5\sigma$ separation from the primary cut for the \lya absorber
  and \lya emitter subsets, respectively.  This corresponds to
  $\gtrsim2.5\sigma$ from each of the respective ($G-\cal{R}$) and
  $\cal{R}$ distribution mean values.  Objects beyond these cuts
  produce $\gtrsim90$\% pure samples of either spectral type.  Objects
  fainter than $\cal{R}$ $=25.5$ that lay $\ge3\sigma$ from the aLBG
  color distribution mean value comprise an expected population of
  LAEs.  This region is shown bounded by dotted lines.}
\label{1and4}
\end{center}
\end{figure}

We acquired the spectroscopic data using the Low-Resolution Imaging
Spectrometer \citep[LRIS;][]{o95,mccarthy98} mounted on the Keck I
telescope on 2009 July 20 using the $400/3400$ grism on the new blue
arm and $400/8500$ grating on the new red arm yielding a blue/red
resolution of $\sim300$ km s$^{-1}$.  Time constraints resulted in a
total integration time of $3600$s with $\sim0.''9$ seeing FWHM.  This
is a $\sim0.5\times$ the total integration time of typical \zzz LBG
spectroscopy, but is found to be sufficient to identify most $\cal{R}$
$\gtrsim25.5$ candidates and the \lya emission feature of $\cal{R}$
$\gtrsim27.0$ LAEs.

\begin{figure*}
\begin{center}
\scalebox{0.30}[0.25]{\rotatebox{90}{\includegraphics{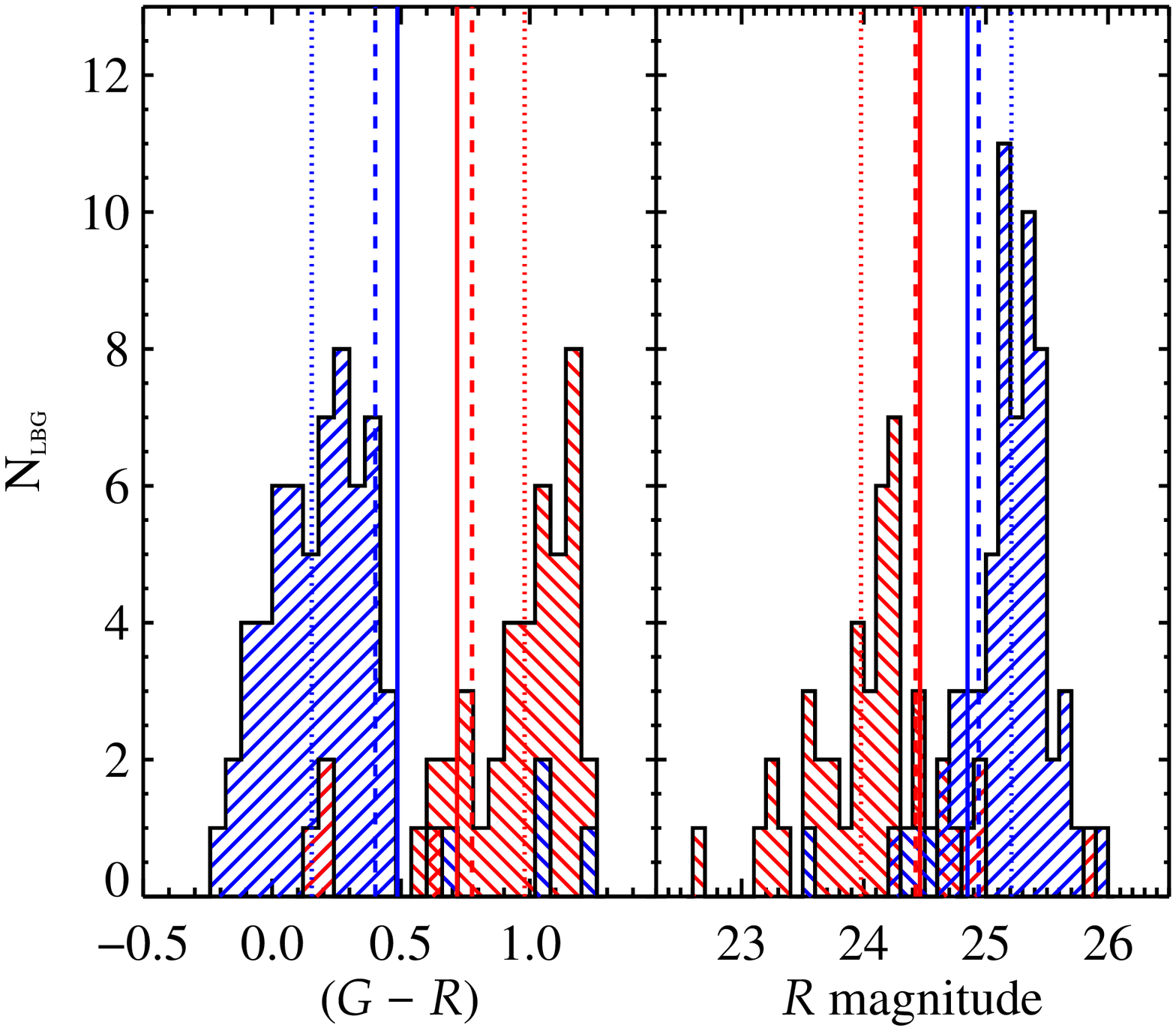}}}
\scalebox{0.30}[0.25]{\rotatebox{90}{\includegraphics{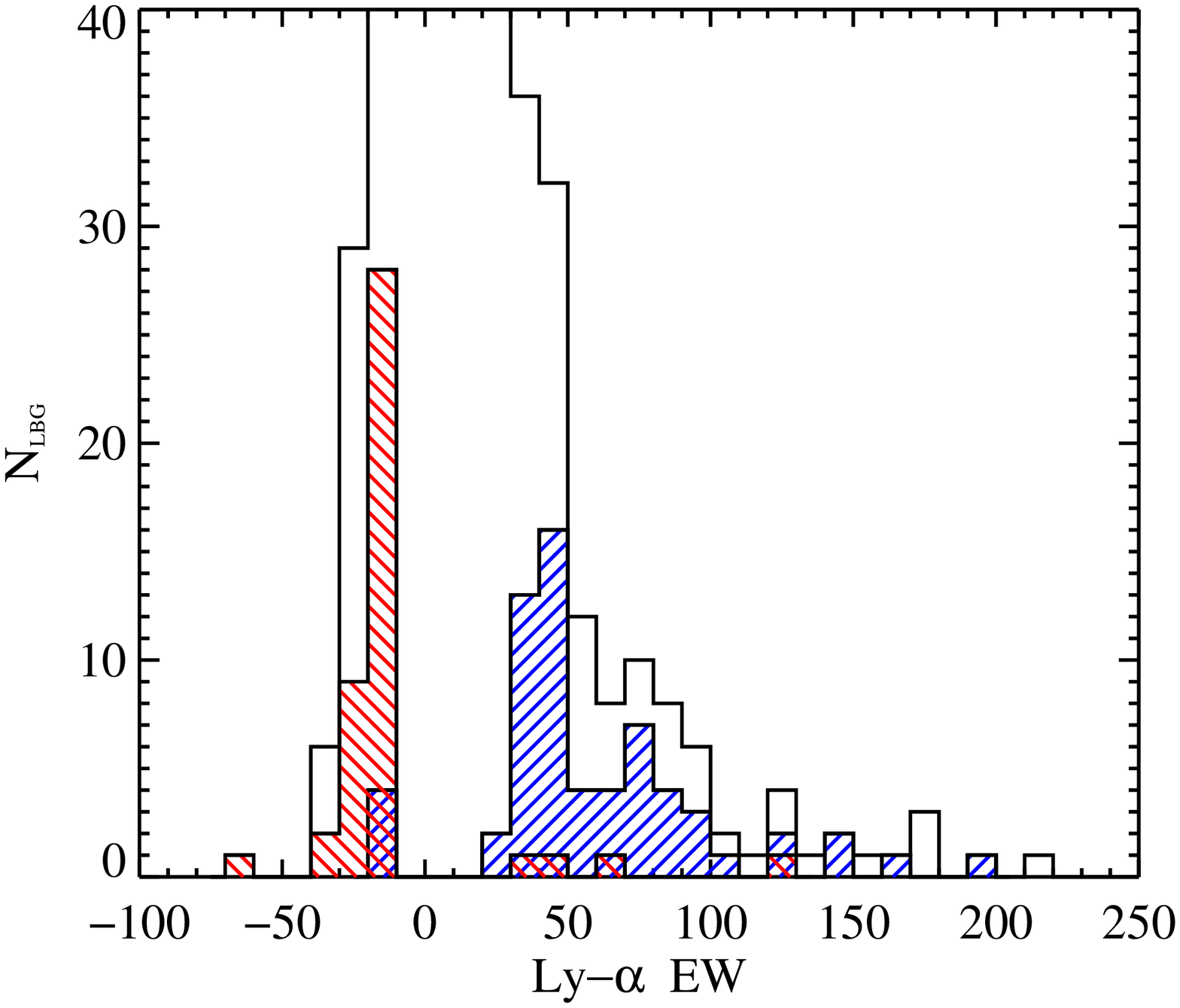}}}
\scalebox{0.30}[0.25]{\rotatebox{90}{\includegraphics{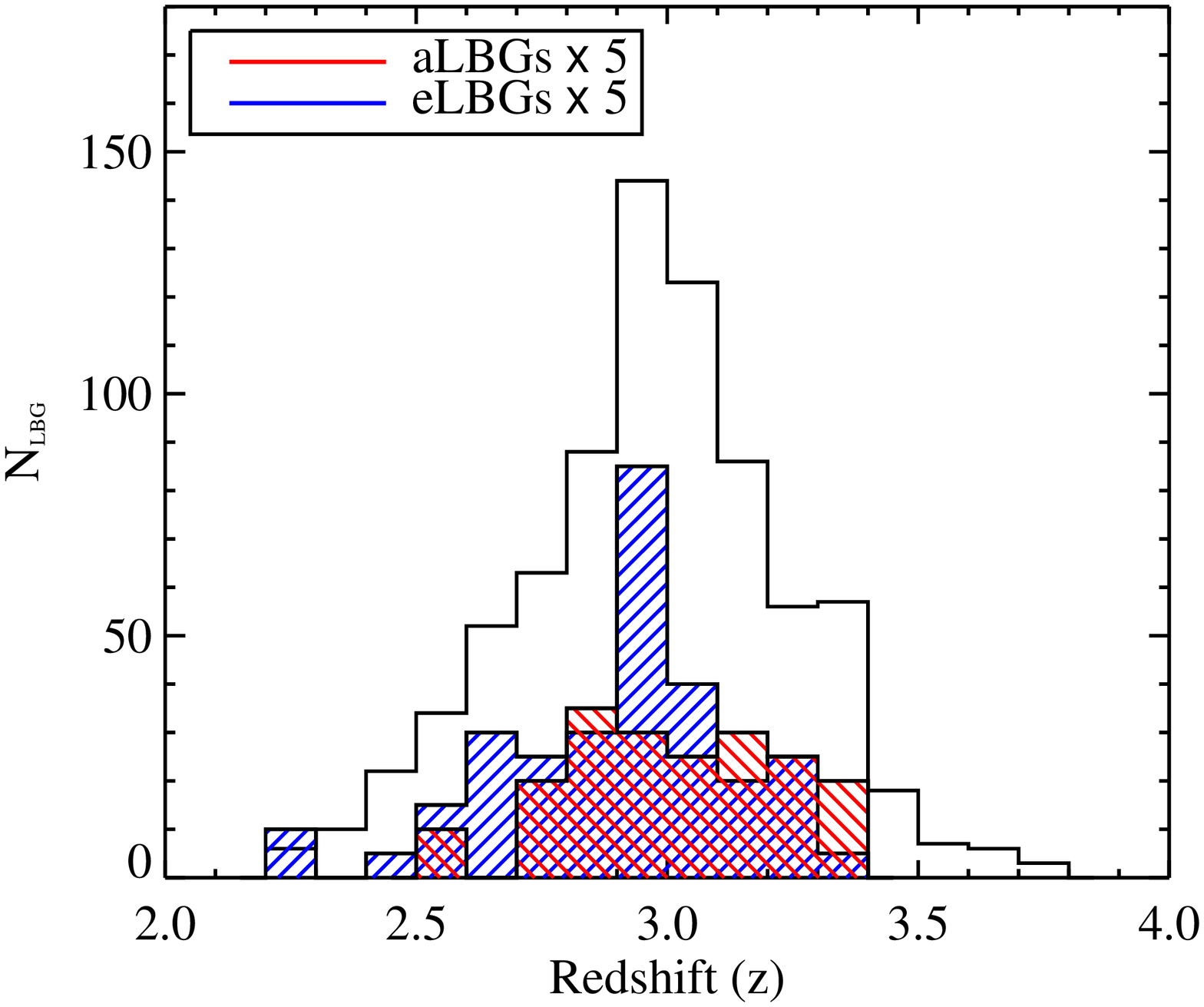}}}
\caption
{\small Histograms of the isolated spectral-type samples obtained when
  applying Equations~\ref{aLBG} and~\ref{eLBG} to the spectroscopic
  dataset of \citet{s03}.  Top: the back-hatched and forward-hatched
  histograms indicate the $\gtrsim90$\% pure aLBG and eLBG isolated
  spectral-type samples, respectively.  In each sample gray (red)
  denotes aLBGs and black (blue) denotes eLBGs.  Vertical lines mark
  the mean values of the aLBG/eLBG ((gray/red; black/blue) solid),
  subset 1/subset 2 ((gray/red; black/blue) dashed), and spectral-type
  sample [(gray/red; black/blue) dot-dash] distributions.  Center: the
  net \lya EW distribution for the total spectroscopic data set
  (unfilled; see \citet{aes03} for the full-scale histogram) and the
  isolated spectral-type samples.  Bottom: the redshift distribution
  of the full sample (unfilled) and the isolated spectral-type samples
  scaled by a factor of 5 to better compare the form of the
  distributions (i.e., any difference between the isolated samples and
  the isolated samples and the full sample is amplified by a factor of
  5).}
\label{cuthist}
\end{center}
\end{figure*}

\section{ANALYSIS}\label{analysis}

We construct a ($G - \cal{R}$) versus ${\cal{R}}$ color--magnitude
diagram (CMD) for the CCS03 LBG photometric sample, plotted in the
left panel of Figure~\ref{color}.  Motivated by the results of
\citet{c09a}, we overlay the spectroscopically confirmed LBGs and code
them by their AES03 defined \lya EW quartiles.  AES03 find that LBGs
with strong \lya have blue $\sim1300-2000$\AA~continua with a
continuous reddening trend seen with decreasing \lya emission EW,
through to strong absorption.  This is the main cause of the color
separation on the CMD in Figure~\ref{color}.  The presence of the \lya
feature in the $G$ band also contributes to the color separation, but
to a lesser extent, and is on order of the $\lesssim0.2$ mag
uncertainties.

We use the term ``aLBGs '' for LBGs with dominant \lya in absorption
(net \lya EW $<0$) and ``eLBGs'' for dominant \lya in emission (net
\lya EW $>0$).  We find that the aLBG and eLBG distributions show
significant overlap but are distinct populations.  A two-sided K--S
test of the aLBG and eLBG ($G-\cal{R}$) distributions produces a
probability of $p=1.5\times10^{-13}$ that the two distributions are
pulled from the same population.  Similarly, a K--S test produces
$p=3.0\times10^{-13}$ for the $\cal{R}$ mag distributions.  This is
illustrated in the right panels of Figure~\ref{color}.

Next, we study two LBG spectral subsets that exhibit the opposite
extremes in \lya EW behavior.  Subset 1 contains the $150$ strongest
aLBGs (net \lya EW $\le-12.0$) and subset 2 consists of the $150$
strongest eLBGs (net EW $\ge26.5$) of the full spectroscopic sample.
The ($G-\cal{R}$) versus $\cal{R}$ CMD of the two subsets is plotted
in Figure~\ref{1and4} and shows two cohesive distributions with a more
distinct separation when compared to the aLBG/eLBG distribution values
shown in Figure~\ref{color}.  Using the statistical mean values of the
two subsets, we sever the populations in both color and magnitude as
shown by the solid line in Figure~\ref{1and4}.  This line falls
$\sim1\sigma$ from both the color and magnitude mean values of both
distributions and defines our primary cut.  We show that more
restrictive selections based on the slope of this cut (defined below),
are very effective in isolating LBG spectral types with differing \lya
and continuum profiles.

\subsection{Spectral-Type Selection Criteria}\label{criteria}

We use the parameters of the subset 1 and 2 distributions to refine
the primary cut in an effort to generate pure samples of each
spectral type.  Choosing CMD cuts that are separated from the primary
cut by $1.5\sigma$ of the ($G-\cal{R}$) distributions of each subset
and along the same slope as the primary cut, selects LBG spectral
types $\gtrsim2.5\sigma$ from the mean values of the opposite
distribution.  We find that the following criteria select nearly pure
samples of aLBGs:
\begin{equation}\label{aLBG}
(G - R)~ \ge ~0.4047~R - 9.376 + 1.5\sigma_{A},
\end{equation} 

\noindent and eLBGs
\begin{equation}\label{eLBG}
(G - R)~ \le ~0.4047~R - 9.376 - 1.5\sigma_{E},
\end{equation}

\noindent where $\sigma_{A}=0.3095$ and $\sigma_{E}=0.2392$ and are
the $1\sigma$ standard deviations of the ($G-\cal{R}$) distributions
for subset 1 and subset 2, respectively.  These selection cuts are
indicated in Figure~\ref{1and4}.

Applying Equations~\ref{aLBG} and~\ref{eLBG} to the full spectroscopic
data set of CCS03 produces spectral-type samples consisting of $40$
aLBGs and $60$ eLBGs with contamination fractions of $0.100$ and
$0.067$, respectively (Figure~\ref{cuthist}; top panel).  In addition,
Figure~\ref{cuthist} presents the net \lya EW and redshift
distributions for the two isolated samples (center and bottom panels,
respectively).  The samples exhibit $66\pm36$ \AA~EW for the eLBGs and
$-20\pm10$ \AA~EW for the aLBGs and their redshift distributions are
found to be representative of the full data set ($z_{FULL}$ =
$2.97\pm0.27$, $z_{aLBG}$ = $3.05\pm0.25$, and $z_{eLBG}$ =
$2.88\pm0.24$).  The above criteria, of course, can be made more
strict by using a larger coefficient of $\sigma_{A}$ and $\sigma_{E}$,
thereby producing samples that are more pure, at the cost of sample
size.  In this manner, large photometric data sets can obtain very
clean spectral-type samples while still maintaining a large number of
objects for good statistics.

\subsection{\lya emitters}\label{laes}

LAEs at \zzz have colors similar to eLBGs with the bulk expected to
have ($G - \cal{R}$) $\lesssim0.0$ \citep[see][Appendix A]{reddy09}.
The colors and mass of LAEs \citep[e.g.,][]{gaw07,lai08} appear to
provide a natural extension of the LBG population and would help to
complete the $\cal{R}$$\sim25.5$ eLBG magnitude truncation.  

From the aLBG/eLBG color and magnitude distributions and those of
their spectral subsets, and the results of the spectral-type selection
criteria above, we find that a pure sample of LAEs (defined here as
having $\cal{R}$ $>25.5$) can be obtained using
\begin{equation}\label{LAE}
(G - R)~ \le ~0.4047~R - 9.376 - 2.0\sigma_{E} \mbox{~and~} R \ge 25.5,
\end{equation}

\noindent which modifies the eLBG cut to a larger separation from the
primary cut.  This effectively avoids systems displaying \lya in
absorption by selecting objects $\gtrsim3\sigma$ from the mean of the
aLBG distribution.  The region of the CMD defined by
Equation~\ref{LAE} is indicated in Figure~\ref{1and4} and shows that
the few spectroscopically identified objects in the CCS03 sample that
meet these criteria exhibit \lya in emission.  Similar to above, the
purity of the LAE sample can be determined by the $\sigma_{E}$
coefficient.  

\subsection{Spectroscopic Tests of the Spectral-Type Predictions}
\label{test}

\begin{figure}
\begin{center}
\scalebox{0.35}[0.35]{\rotatebox{90}{\includegraphics{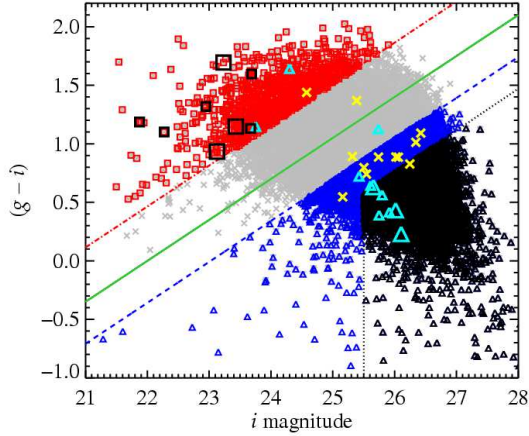}}}
\caption
{\small ($g -i$) vs. $i$-band CMD for $u^*gi$ selected \zzz objects in
  the CFHTLS Deep field ``D4'' (see the text).  Objects are plotted
  similar to Figure~\ref{1and4} but use light gray (black) triangles
  to indicate those meeting the adopted LAE criteria.  A sample of
  $32$ systems were targeted for spectroscopic follow-up.  Black
  squares mark confirmed aLBGs, black (cyan) triangles mark confirmed
  eLBGs, and black (yellow) crosses indicate low S/N unidentified
  systems.  Large squares indicate strong absorption and large
  triangles indicate strong emission.  The images are complete to
  $i\sim 27$ and illustrate the large numbers of each population
  accessible by this method. }
\label{cfht}
\end{center}
\end{figure}

We use $34316$ $u^*gi$-selected \zzz LBGs in the CFHTLS Deep field
``D4'' to test the predictions of the spectral-type criteria and our
LAE assumptions.  Figure~\ref{cfht} plots the ($g - i$) versus $i$
mag CMD for the ``D4'' field and shows our tentative spectral-type
cuts based on the $r<25.5$ \zzz LBG densities and the results of the
above analysis.  The primary cut shown here produces aLBG/eLBG samples
with a very similar $r\le25.5$ ratio ($\sim4600/5200$) as compared to
the $\cal{R}$ ratio of CCS03 when maintaining conventional
color-selection criteria, and specifically the constraint ($g -i$)
$\ge-0.2$.
For more complete color-space detection, we relax the criterion to ($g
- i$) $\ge-1.0$ to include the small number of bluer objects to probe
the full eLBG and expected LAE ($g -i$) distribution.  The relaxation
of the color in this manner does not increase the fraction of
interlopers \citep[e.g., CCS03;][]{c05}.  The ``D4'' field is
reflective of the remaining three CFHTLS Deep fields in depth and
generates spectral-type samples of $\sim1600$ aLBGs and $\sim14,000$
eLBGs, $\sim8000$ of which fall in the LAE selection region (LAE
magnitude definition shown in Figure~\ref{cfht} is $i>25.5$).

The analysis of our $32$ spectra (Section~\ref{data}) indicates that
our tentative spectral-type cuts are very effective.  Of the $11$ LBGs
targeted in the aLBG spectral-type cut, eight have confirmed \lya in
absorption, two show complex absorption and emission profiles with net
\lya emission, and one is unidentified as a result of its low S/N.
Nine of the 17 LBGs targeted in the eLBG region were unidentified
(although six show weak evidence of emission), with eight exhibiting
\lya emission.  Seven of the eight galaxies that fall in the LAE
region show \lya emission, with the remaining galaxy being
unidentified.  This demonstrates a high efficiency in identifying
$i\lesssim27.0$ LAEs from their broadband colors and helps to confirm
the LAE extension to the LBG population.


\section{CONCLUSION}\label{disc}

We present an analysis of the photometric properties of the \zzz LBG
spectroscopic sample of \citet{s03}.  The relationships between \lya
EW and ($G-\cal{R}$) and \lya EW and $\cal{R}$ mag enable a spectral
separation of the LBG population using broadband data.  We define
statistical photometric cuts to reliably generate $\gtrsim90$\% pure
spectral-type samples of LBGs displaying dominant \lya in absorption
and dominant \lya in emission.  In addition, the spectral-type
broadband criteria are extended to isolate clean samples of LAEs.

Our spectroscopic sample of $32$ \zzz LBGs from the CFHTLS
demonstrates the efficiency of the broadband criteria presented here
in identifying galaxies based on their \lya feature, including LAEs.
Use of this method will allow the statistical study of \zzz galaxies
populations from the large numbers easily acquired in deep broadband
surveys.  This circumvents the expense and constraints of
investigations using MOS spectroscopy and the limitations of
narrowband searches.  Further tests of the criteria, and the extension
of the results to other redshifts and filter sets, are presented in a
future paper.


\acknowledgments

J. C. thanks A. E. Shapley for helpful discussions and access to the
spectroscopic values.  J. C. gratefully acknowledges generous support
by Gary McCue.


\end{document}